\documentclass[aps,pra,reprint]{revtex4-1}

\usepackage{xcolor}
\usepackage{tikz}
\usepackage{amsmath}
\usepackage{amssymb,amsmath}
\usepackage{graphicx}
\usepackage{upgreek}
\usepackage[T1]{fontenc}
\usepackage[utf8]{inputenc}
\usepackage{graphicx}
\usepackage{bm}

\begin{document}

\title{Superscattering, Superabsorption, and Nonreciprocity in Nonlinear Antennas}
\author{Lin Cheng{$^{1,2}$} Rasoul Alaee,{$^{1,\ast}$} Akbar Safari,{$^{1}$} Mohammad Karimi,{$^{1}$} Lei Zhang,{$^{2}$} and Robert W. Boyd{$^{1,3}$}}

\address{$^1$Department of Physics, University of Ottawa, Ottawa Q1N 6N5, Canada}
\address{$^{2}$School of Electronic and Engineering, Xi'an Jiaotong University, Xi'an, 710049, China.}
\address{$^{3}$The Institute of Optics, University of Rochester, Rochester, New York 14627, USA.\\
$*$ $\rm{Corresponding\,\, author: rasoul.alaee@gmail.com}$
}


\begin{abstract}
We propose tunable nonlinear antennas based on an epsilon-near-zero material with a large optical nonlinearity. We show that the absorption and scattering cross sections of the antennas can be controlled \textit{dynamically} from a nearly superscatterer to a nearly superabsorber by changing the intensity of the laser beam. Moreover, we demonstrate that a hybrid nonlinear antenna, composed of epsilon-near-zero and high-index dielectric materials, exhibits \textit{nonreciprocal} radiation patterns because of broken spatial inversion symmetry and large optical nonlinearity of the epsilon-near-zero material. By changing the intensity of the laser, the radiation pattern of the antenna can be tuned between a bidirectional and a unidirectional emission known as a Huygens source. Our study provides a novel approach toward ultrafast dynamical control of metamaterials, for applications such as beam steering and optical limiting.\end{abstract}

\maketitle
Optical antennas as fundamental building blocks in nanophotonics and metamaterials allow us to manipulate and control optical fields on the nanometer scale~\cite{Novotny2009}. By localizing the energy of a propagating wave, optical antennas provide enhanced control on light-matter interaction for applications such as microscopy~\cite{Taylor2019} and nonlinear optics~\cite{Krasnok2018}. Scattering and absorption cross sections are the most important quantities to describe how strong an antenna interacts with the incident light. These cross sections depend on the induced electric and magnetic multipole moments~\cite{Alaee2019} and can be tailored by engineering either the geometry or the material properties of the optical antennas. Considering their wide applications in photonics, various optical antennas have been proposed to achieve fascinating scattering phenomena including directional emission known as the Kerker effects~\cite{kerker1983,zambrana2013duality,Alaee:2015generalized}, superscattering~\cite{Ruan2010,Ruan2011}, superabsorption~\cite{Tribelsky:2006,Ng:2009,Miroshnichenko:2018,Estakhri:2014,Rahimzadegan:17}, optical cloaking~\cite{Andrea:2008}, and non-radiating scattering states~\cite{Devaney:1973,Hsu:2014}. 

\begin{figure}
\centering
\includegraphics[width=\columnwidth]{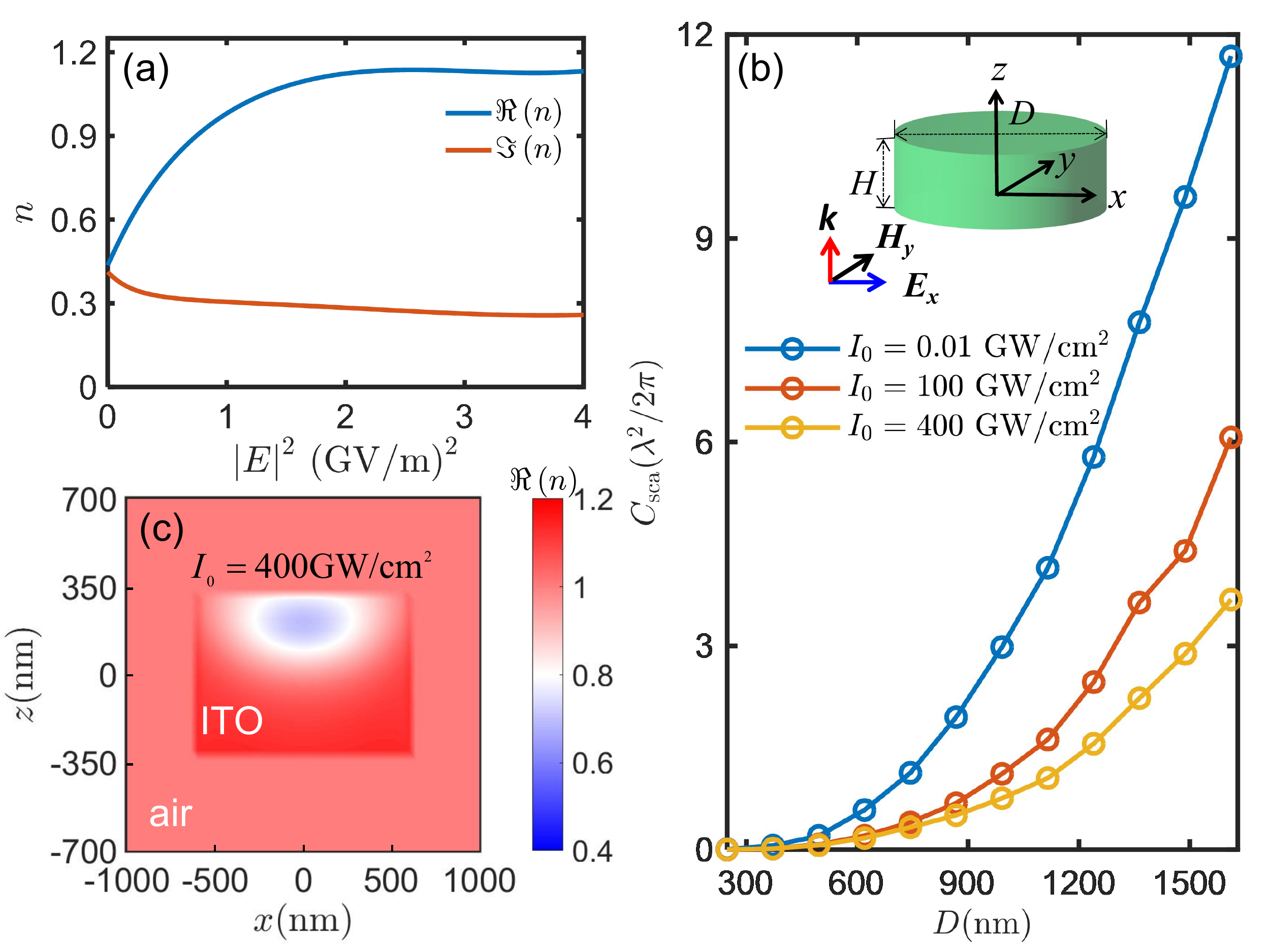}
\caption{\textit{Nonlinear antenna based on epsilon-near-zero (ENZ) material:} (a) The real and imaginary parts of the intensity-dependent refractive index of ITO at the ENZ wavelength $\lambda_{\rm ENZ}=1240$\,nm~(see Ref.~\cite{reshef:2017}). (b) Scattering cross section (normalized to $\lambda^2/2\pi$) of the ITO antenna at the ENZ wavelength as a function of the diameter of the antenna $D$ for different intensities. Note that the lowest intensity, i.e., $I_0 =0.01\mathrm{GW}/\mathrm{cm^2}$, corresponds to the linear response of the antenna. Inset shows a schematic of the ITO disk with height $H$ and diameter $D$. The ITO antenna is illuminated by an $x$-polarized plane wave propagating in the $z$ direction. (c) The real part of refractive index in $xz$-plane (at $y=0$) at the highest intensity, i.e., $I_0=400$ $\mathrm{GW}/\mathrm{cm^2}$. We assume that the height of the ITO antenna is $H=D/2$ and the surrounding medium is air.}
\label{Fig_ITO}
\end{figure}

In order to realize a versatile control of electromagnetic radiation, it is highly desirable to modulate the optical properties of the antennas dynamically. Recently, epsilon-near-zero (ENZ) materials have been shown to exhibit an exceptionally large intensity-dependent refractive index~[see Fig.~\ref{Fig_ITO}(a)]~\cite{Alam:2016}. Therefore, ENZ materials provide a new platform to optically tune the response of the material within a subpicosecond timescale~\cite{Alam:2016}. Specifically, negative refraction, tunable metasurfaces, optical switches, tunable cavities, and coherent perfect absorbers have been achieved using indium tin oxide (ITO) and aluminum-doped zinc oxide (AZO) as ENZ materials~\cite{Argyropoulos:2012,Kinsey:2015AZO,Alam:2016,Caspani:2016,reshef:2017,Liberal:2017,Clerici:2017,Ferrera:2017,Liberal:2017rise,Alam2018nature,Vezzoli:2018,Kim:2018Cavity,Niu:2018Review,Reshef:2019review,Kinsey2019,Kinsey2019,Alaee:2020CPA,Bruno:2020,Paul:2020,Bruno2020:CPA,Bruno:2020shift}.

In this work, we theoretically study the optical response of nonlinear antennas composed of epsilon-near-zero and high-index dielectric materials. Previously, nonlinear antennas have been realized theoretically and experimentally to enhance nonlinear responses such as second-harmonic and third-harmonic generation~\cite{Smirnova2016:multipolar,Camacho2016:nonlinear,Smirnova:2019,Smirnova:2018multipolar,Carletti:2018}. However, the Kerr-type nonlinearity has not been strong enough to significantly modify the scattering response of the antennas at the fundamental wavelength. 
In our work, the large Kerr-type nonlinearity of the ENZ material plays a crucial role in changing the induced multipole moments, and thus drastically modifying its scattering, absorption, and extinction cross sections as well as its radiation pattern. In particular, we can optically switch the antennas' response from a nearly superscattering to a nearly superabsorbing state. By employing the multipole expansion of the induced \textit{intensity-dependent} polarization current, we show that the radiation pattern of the antennas can be tuned from a non-directive to a nearly Huygens source by changing the intensity of the laser beam. 

\textit{Nonlinear antennas based on ENZ materials.—} Let us consider a nonlinear antenna made of indium tin oxide (ITO)~[see the inset of Fig.~\ref{Fig_ITO}(b)]. The antenna is illuminated by an $x$-polarized time-harmonic plane wave with electric field $\mathbf{E_{\rm inc}}\left(\mathbf{r},\omega\right)=(E_0/2)e^{i({\mathbf{k}\cdot{{\mathbf r}}-\omega{t}})}{\bf{e}}_x+{\rm c.c.}$, where $ |{\bf{k}}|=2\pi/\lambda$ is the wavenumber, $\omega$ is the angular frequency, $E_0$ is the amplitude of incident field, and c.c. means complex conjugate. $I_0=\frac{1}{2}c\varepsilon_0|E_0|^2$ is the free-space intensity of the incident beam. The intensity-dependent
refractive index of ITO is $n_{_{\rm NL}}\left(\mathbf{r},\omega\right)=\sqrt{ \varepsilon_{_{\rm NL}}\left(\mathbf{r},\omega\right)}$, where $\varepsilon_{_{\rm NL}}\left(\mathbf{r},\omega\right)$ is given by~\cite{boydbook,reshef:2017}
\begin{equation}
   \varepsilon_{_{\rm NL}}\left(\mathbf{r},\omega\right)\approx
   \varepsilon_{{\rm L}}+
 \overset{3 }{\underset{j=1}{\sum }}c_{2j+1}
   \chi^{(2j+1)}\left(\omega\right)\left|\frac{\mathbf{E}\left(\mathbf{r},\omega\right)}{2}\right|^{2j},\label{n_ITO}
\end{equation}
where $\chi^{(3)}\left(\omega\right)$, $\chi^{(5)}\left(\omega\right)$, and  $\chi^{(7)}\left(\omega\right)$ are the third-order, fifth-order, and seventh-order nonlinear susceptibilities (see Table 1 in Ref.~\cite{reshef:2017}) of ITO, respectively. $c_3=3,\,c_5=10,\,c_7=35$ are the degeneracy factors~\cite{boydbook} and $\mathbf{E}\left(\mathbf{r},\omega\right)$ is the electric field inside the ITO.

The real part of the permittivity of ITO is zero at $\lambda_{\rm ENZ}=1240$\,nm, which is called the ENZ wavelength. As a consequence, ITO exhibits a large nonlinear refractive index around its ENZ wavelength~\cite{Alam:2016,Caspani:2016,reshef:2017}. 
Figure~\ref{Fig_ITO}(a) plots the real and imaginary parts of the intensity-dependent refractive index of ITO film at $\lambda_{\rm ENZ}$~\cite{Alam:2016,reshef:2017}. Note that the change in the real part of the refractive index by intensity is approximately 0.72, which is even larger than the linear refractive index of ITO, 0.4. In the following, we incorporate this large nonlinear response of ITO at the ENZ wavelength, and perform our simulations using a Maxwell’s equations numerical solver combined with an iterative method to solve for an intensity-dependent refractive index inside the antenna. 

In order to understand the optical response of the proposed nonlinear antennas in this Letter, we employ multipole expansion of the induced nonlinear (intensity-dependent) polarization current $\mathbf{J}_{_{\rm NL}}\left(\mathbf{r},\omega\right)=-i\omega\left[\varepsilon_{_{\rm NL}}\left(\mathbf{r},\omega\right)-\varepsilon_{0}\right]\mathbf{E}\left(\mathbf{r},\omega\right)$~\cite{Alaee:2018,Alaee2019}. Through use of the induced multipole moments the total scattering cross section of the nonlinear ITO antenna (sum of the induced multipole moments) can be calculated by~\cite{Alaee:2018,Alaee2019}
\begin{eqnarray}
C_{{\rm sca}} & = & \frac{k^{4}}{6\pi\varepsilon_{0}^{2}|E_{0}|^{2}}\left[\sum_{\alpha}\left(\left|p_{\alpha}\right|^{2}+\left|\frac{m_{\alpha}}{c}\right|^{2}\right)\right.\nonumber\\
 &  & \left.+\sum_{\alpha,\beta}\left(\left|kQ_{\alpha\beta}^{e}\right|^{2}+\left|\frac{kQ_{\alpha\beta}^{m}}{c}\right|^{2}\right)\right],\label{Csca}
\end{eqnarray}
where, $\alpha,\beta=x,y,z$, and where $p_{\alpha}$, $m_{\alpha}$, $Q_{\alpha\beta}^e$, and $Q_{\alpha\beta}^m$ are the
electric dipole (ED), magnetic dipole (MD), electric quadrupole (EQ), and magnetic quadrupole (MQ) multipole moments.

Figure~\ref{Fig_ITO}(b) shows the scattering cross section (normalized to $\lambda^2/2\pi$) of the ITO disk as a function of its diameter $D$ for three different intensities. It can be seen that the scattering cross section gradually increases as $D$ increases. At high intensities, e.g., $I_0=400$ $\rm{GW/cm^2}$, the scattering cross section is approximately 4 times smaller than the linear response, i.e., $I_0=0.01$ $\rm{GW/cm^2}$~[see the Supplementary Material (SM)~\cite{SM}. At high intensities, the refractive index of ITO is close to the refractive index of the surrounding medium (air), and thus scattering becomes smaller than that of low intensities~[Fig.~\ref{Fig_ITO}(a) and (c)]. This results can also be understood in terms of the induced electric and magnetic multipole moments~(see SM~\cite{SM} for details). In Fig.~\ref{Fig_ITO}(c), we plot the real part of the refractive index of the ITO antenna in $xz$-plane. The refractive index depends on position $\mathbf{r}$ because of the induced \textit{nonuniform} electric field distribution inside the antenna, i.e., ${\bf{E}}({\bf{r}})$~[see also Eq.~(\ref{n_ITO})].

\begin{figure}
 \centering
 \includegraphics[width=\columnwidth]{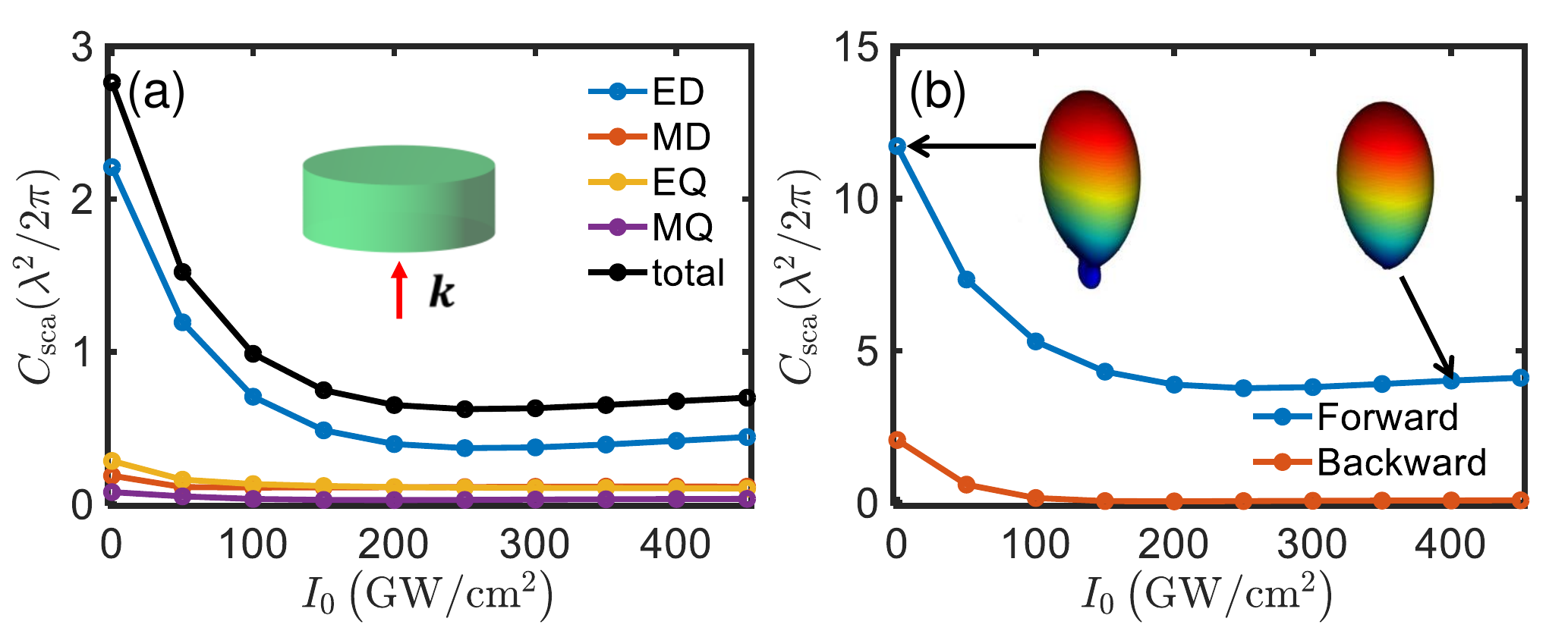}
\caption{\textit{Intensity-dependent response of nonlinear antenna:} (a) Total scattering cross section (normalized to $\lambda^2/2\pi$) and contribution of different electric and magnetic multipole moments as a function of the input intensity for the ITO antenna. Contribution of different multipoles are labeled by ED (electric dipole), MD (magnetic dipole), EQ (electric quadrupole), and MQ (magnetic quadrupole). The geometrical parameters of the ITO antenna are $D=0.8~\lambda_{\rm ENZ}$ and $H=D/2$. (b) Normalized forward and backward scattering cross sections calculated from Eq.~(\ref{Csca_FB}). The insets illustrate normalized far-field radiation patterns of the ITO antenna for low and high intensities.}
\label{Fig_ITO_radiation}
\end{figure}

 Figure~\ref{Fig_ITO_radiation}(a) shows the total scattering cross section and contributions of the electric and magnetic multipole moments as a function of intensity. Although the contribution of the ED moment is significantly larger than any other modes, the antenna radiates mainly in the forward direction because of the presence of higher order moments~[Fig.~\ref{Fig_ITO_radiation}(a)-\ref{Fig_ITO_radiation}(b)]. The normalized forward $C_{\mathrm{sca}}^{\mathrm{F}}$ and backward $C_{\mathrm{sca}}^{\mathrm{B}}$ scattering cross sections are plotted in Fig. 2(b), which are calculated from~\cite{Alaee:2015generalized}
\begin{equation}\label{Csca_FB}
\begin{aligned}
C_{\mathrm{sca}}^{\mathrm{F/B}}=C_{\mathrm{norm}}\left|p_{x}\pm\frac{m_{y}}{c}\mp\frac{ikQ_{xz}^{e}}{6}-\frac{ikQ_{yz}^{m}}{6c}\right|^{2},
\end{aligned}
\end{equation}
where $C_{\mathrm{norm}}=k^{4}/\left(4\pi\varepsilon_{0}^{2}\left|E_{0}\right|^{2}\right)$. In Eq.~(\ref{Csca_FB}), the forward and backward scattering cross sections are calculated at $\phi_{\rm F/B}=0$ and $\theta_{\rm F/B}=0,\pi$. Note that the ED (and similarly MQ) moment exhibit in-phase forward and backward scattered electric fields, whereas the MD (and similarly EQ) moment show out-of-phase fields~[see $\pm$ in Eq.~(\ref{Csca_FB})]. As the intensity of the laser increases, the contribution of different multipoles changes. Consequently, the antenna radiates solely in the forward direction at high intensities, which is clearly evident from the inset of Fig.~\ref{Fig_ITO_radiation} (b).

\begin{figure}
 \centering
\includegraphics[width=\columnwidth]{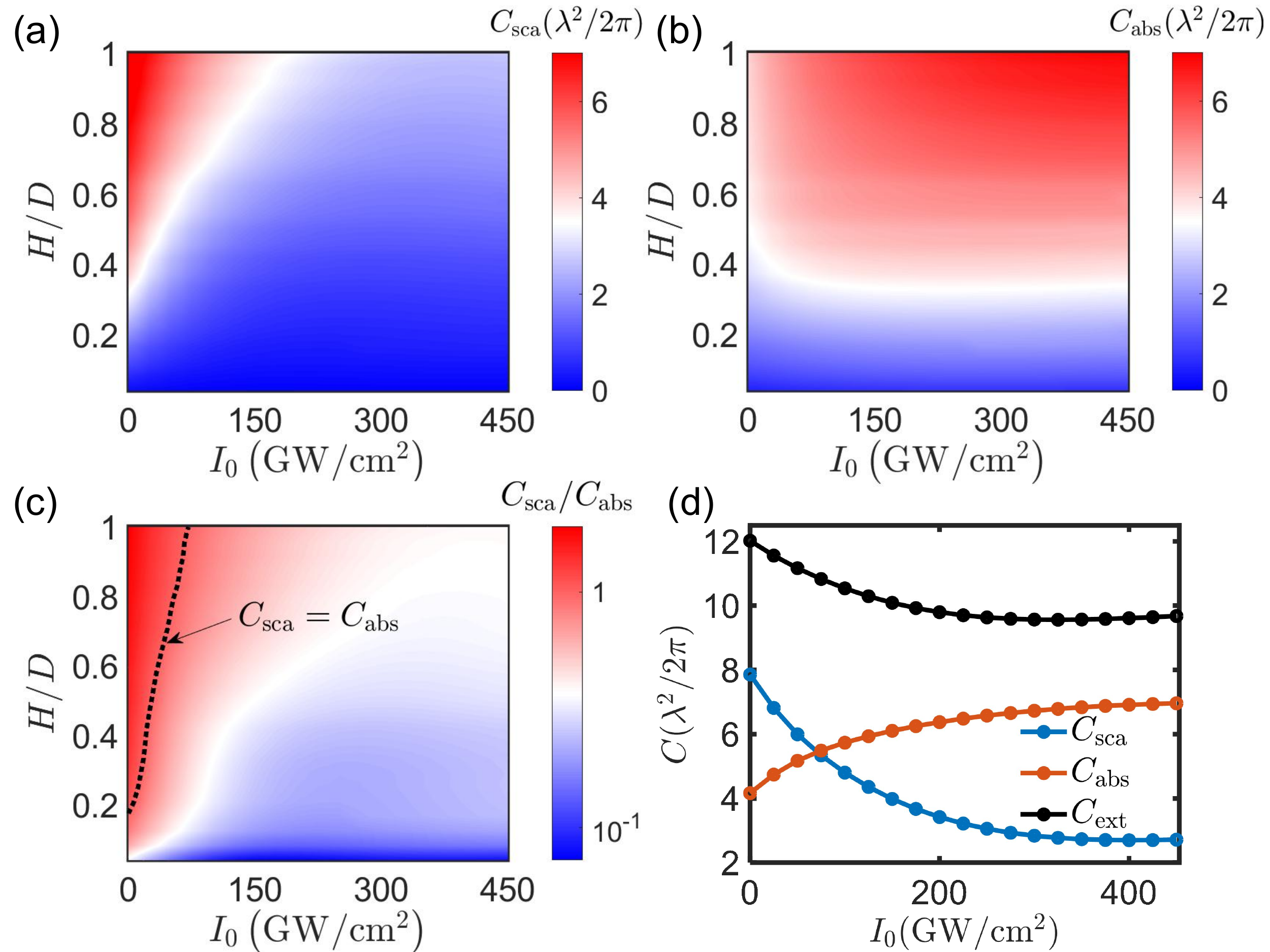}
\caption{\textit{Tunable superabsorption and superscattering based on an ITO antenna:} (a) Total scattering cross section $C_{\rm sca}$ (normalized to $\lambda^2/2\pi$) of the ITO antenna [see the inset of Fig.~\ref{Fig_ITO} (b)] as a function of height-to-diameter ratios, $H/D$, and the intensity of laser beam $I_0$, where the diameter of the antenna $D=1200$\,nm. (b) Same as (a) for the absorption cross section, i.e., $C_{\rm abs}$. (c) The ratio between the scattering and absorption cross section shown in logarithmic scale. The black dashed line indicates the condition that the scattering and absorption cross sections are equal, i.e., $C_{\rm sca}=C_{\rm abs}$ . (d) Scattering, absorption and extinction cross sections as functions of intensity $I_0$ for $H=D=1200$\,nm.}
\label{Fig_ITO_Superscattering}
\end{figure}
\textit{Superabsorption and superscattering in nonlinear antennas.—} The maximum scattering cross section of an isotropic nanoparticle with multipolar response is $C_{{\rm sca,}\,j}^{\rm max}=(2j+1)\lambda^2/2\pi$, where $j$ is the order of the multipole; e.g., $j=1$, $2$, and $3$, for dipoles, quadrupoles, and octupoles, respectively~\cite{Tribelskii:1984,Ruan2010,Ruan2011,Rahimzadegan:17}. For each multipolar order, the maximum scattering occurs at the resonance for a particle with negligible Ohmic losses compared to the radiation loss (this condition is known as \textit{overcoupling})~\cite{Ruan2010,Alaee:2017Review}. By engineering subwavelength nanoparticles, it is possible to achieve a much larger scattering cross section compared to a dipolar one, i.e., $C_{{\rm sca},\,1}^{\rm max}=3\lambda^2/2\pi$. This phenomenon is known as superscattering and is achieved by overlapping resonant frequencies of different multipoles~\cite{Ruan2010,Ruan2011,Estakhri:2014,Rahimzadegan:17}. It has also been shown that the maximum absorption cross section of a nanoparticle with multipolar response is limited to $C_{{\rm abs,}\,j}^{\rm max}=C_{{\rm sca,}\,j}^{\rm max}/4=(2j+1)\lambda^2/8\pi$~\cite{Tribelsky:2006,Miroshnichenko:2018,Estakhri:2014,Rahimzadegan:17}. The enhanced absorption can be achieved for particles that operate at \textit{critical coupling} (i.e., nonradiative or Ohmic loss is equal to the radiation loss).  

Here, we show that scattering and absorption cross sections of an ITO antenna can be tuned between nearly superscattering and nearly superabsorbing states by simply changing the intensity of the input laser beam. Figure~\ref{Fig_ITO_Superscattering} (a)-\ref{Fig_ITO_Superscattering} (b) shows scattering and absorption cross sections of our ITO antenna as a function of the laser intensity for different height-to-diameter ratios, $H/D$. For low intensities and $H/D=1$, the scattering cross section is $C_{\rm sca}\approx2.7\times 3\lambda^2/2\pi$ which is 2.7 times larger than the maximum scattering of a dipole. By increasing the laser intensity, the absorption cross section increases and reaches to $C_{\rm abs}\approx9.3 \times 3\lambda^2/8\pi$ which is significantly larger than that of a dipolar scatterer. Therefore, the ITO antenna behaves as a superscatterer at low intensities and as a superabsorber at high intensities. This behavior can be seen clearly from Fig.~\ref{Fig_ITO_Superscattering} (d) which plots the scattering,  absorption and extinction cross sections as a function of intensity for $H = D = 1200$\,nm. The ratio of the scattering to the absorption cross section is depicted in Fig.~\ref{Fig_ITO_Superscattering} (c). Three distinct coupling regimes are evident: (i) a large scattering cross section compared to absorption, i.e., $C_{\rm sca}\gg C_{\rm abs}$, the area to the left of the black dashed line, (ii) scattering is identical to absorption cross section, i.e., $C_{\rm sca}=C_{\rm abs}$, as indicated by the black, dashed line in Fig.~\ref{Fig_ITO_Superscattering}(c), and (iii) a large absorption cross section compared to scattering, i.e., $C_{\rm abs}\gg C_{\rm sca}$, the area to the right of the black dashed line.


\begin{figure}
 \centering
\includegraphics[width=0.48\textwidth]{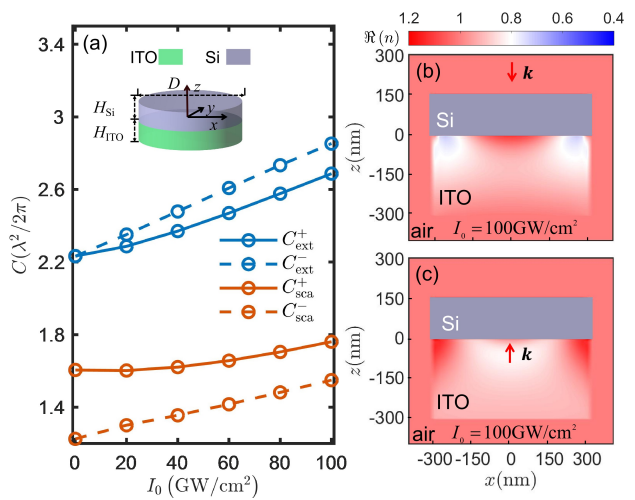}
\caption{\textit{Hybrid nonlinear antenna composed of ITO and silicon disks:} (a) Total extinction ($C_{\rm ext}^{\pm}$) and scattering  ($C_{\rm sca}^{\pm}$) cross sections (normalized to $\lambda^2/2\pi$) of the hybrid antenna as a function of the laser intensity when illuminated by an $x$-polarized plane wave propagating in two opposite directions, i.e., ${\bf{k}}=\pm k_0{\bf{e}}_{z}$. Inset shows the schematic of the hybrid nonlinear antenna. (b) and (c)
The real part of refractive index in $xz$-plane (at $y=0$) at intensity $I_0=100$ $\mathrm{GW}/\mathrm{cm^2}$ for top (${\bf{k}}=-k_0{\bf{e}}_{z}$) and bottom (${\bf{k}}=+k_0{\bf{e}}_{z}$) illuminations, respectively. The surrounding medium is air. The geometrical parameters of the ITO and silicon disks are $D_{\rm ITO}=D_{\rm Si}=620$\,nm, $H_{\rm ITO}=D_{\rm ITO}/2$, and  $H_{\rm Si}=150$\,nm.  }
\label{Fig_hybrid}
\end{figure}
\textit{Hybrid nonreciprocal nonlinear antennas.—} Lossless high-index dielectric antennas support \textit{strong} electric and magnetic responses with large scattering cross sections~\cite{Bohren2008,Evlyukhin:2012,Kuznetsov:2012}. Thus, to achieve an even greater control on the scattering properties of ITO antennas, one can devise a nonlinear antenna composed of ENZ and high-index dielectric materials. In the following, we consider a hybrid nonlinear antenna made of subwavelength ITO and silicon disks, as shown in the inset of Fig.~\ref{Fig_hybrid}(a). Due to the broken inversion symmetry, the hybrid antenna exhibits \textit{magneto-electric} coupling or the so-called bianisotropic response~\cite{Tretyakov:03,Alaee:15D,Asadchy2018bianisotropic}. 


According to the optical theorem and the Lorentz reciprocity, the extinction cross section of an arbitrarily shaped antenna made of reciprocal materials is the same for two opposite illumination directions~\cite{newton:1976,Harrington:2001,Sounas:2014Extinction,Alaee:15D}. However, the scattering and absorption cross sections of the reciprocal antenna depend on the illumination direction due to absorption (Ohmic) losses which are related to the induced bianisotropic response~\cite{,Sounas:2014Extinction,Alaee:15D}. In Fig.~\ref{Fig_hybrid}(a) we plot the extinction and scattering cross sections of the hybrid antenna when illuminated from opposite directions (${\bf{k}}=\pm k_0{\bf{e}}_{z}$). In the linear regime (low intensities) the antenna is reciprocal. Therefore, the extinction cross section for top and bottom illuminations are identical, i.e., $C^{+}_{\rm ext}=C^{-}_{\rm ext}$~\cite{Sounas:2014Extinction,Alaee:15D}. However, the scattering (and also the absorption) cross sections of the antenna are different, $C^{+}_{\rm sca}\neq C^{-}_{\rm sca}$. 

\begin{figure}
\centering
\includegraphics[width=0.48\textwidth]{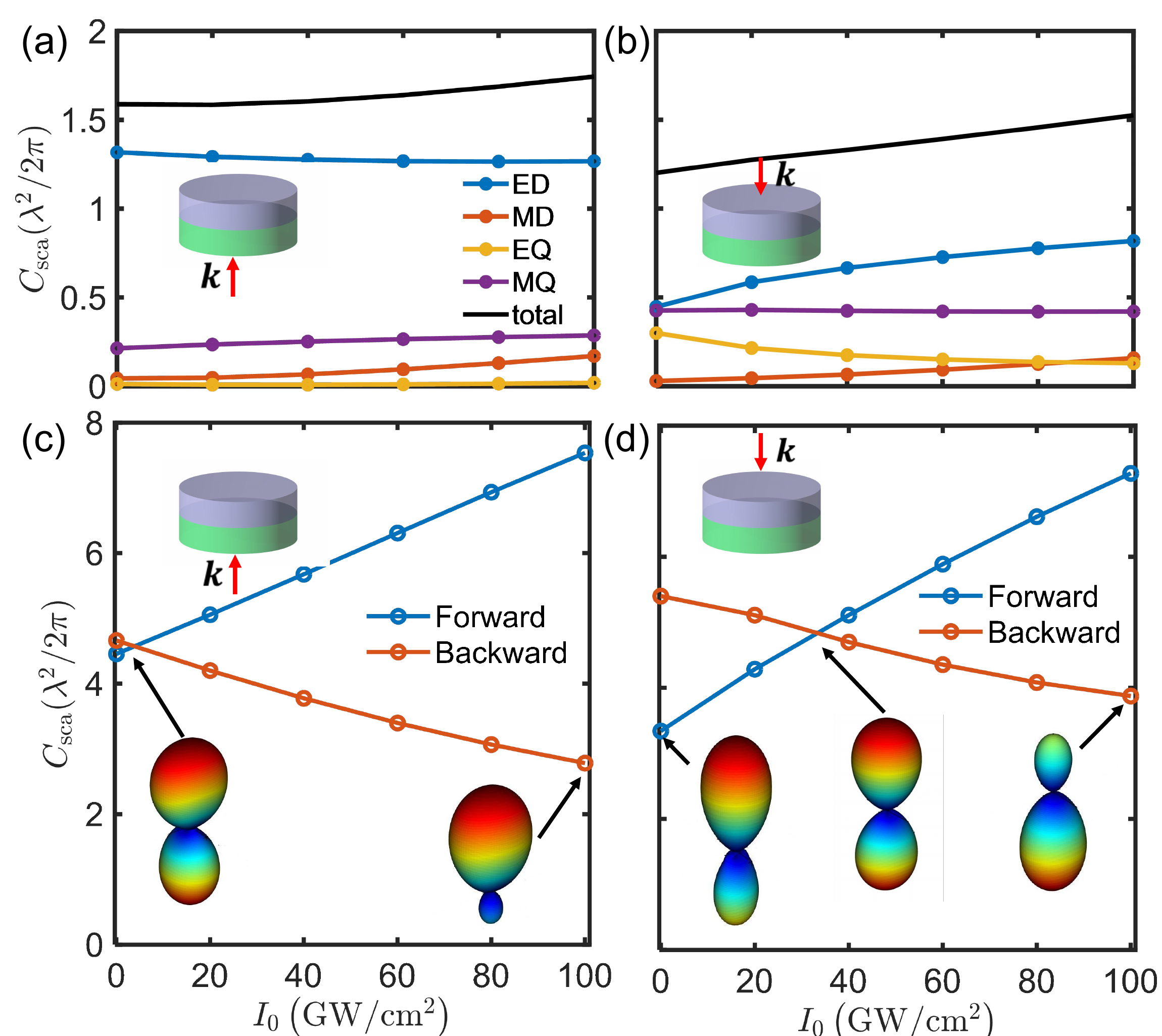}
\caption{\textit{Scattering and radiation patterns of hybrid nonlinear antennas:} (a) Total scattering cross section (normalized to $\lambda^2/2\pi$) and  contribution of different electric and magnetic multipole moments of the hybrid antenna as a function of the laser intensity for the bottom illumination direction, i.e., ${\bf{k}}=+ k_0{\bf{e}}_{z}$. 
(c) Normalized forward (blue line) and backward (red line) scattering cross sections calculated from Eq.~(\ref{Csca_FB}) for the bottom illumination direction. Insets illustrate far-field radiation patterns of hybrid nonlinear antenna in low ($I_0=0.01$ $\mathrm{GW}/\mathrm{cm^2}$) and high ($I_0=100$ $\mathrm{GW}/\mathrm{cm^2}$) intensities for the bottom illumination direction, i.e., ${\bf{k}}=+ k_0{\bf{e}}_{z}$. (b) and (d) Same as (a) and (c) for the top illumination direction, i.e., ${\bf{k}}=-k_0{\bf{e}}_{z}$.
}
\label{Fig_hybrid_radiation}
\end{figure}

At high laser intensities, the hybrid antenna is \textit{nonreciprocal} and two conditions to break the Lorentz reciprocity are simultaneously satisfied, i.e., the large optical nonlinearity and lack of inversion symmetry~\cite{Asadchy:2020}. Thus, as shown in Fig.~\ref{Fig_hybrid}(a), the extinction cross sections of the antenna are not the same for the two opposite illuminations at high intensities. Figure~\ref{Fig_hybrid}(b) and \ref{Fig_hybrid}(c) shows the real part of the refractive index in the $xz$-plane for $I_0=100\,\rm{GW/cm^2}$ and indicate a position-dependent refractive index. Clearly, the refractive indices are different for opposite illuminations: $\ensuremath{n_{_{\rm NL}}^{+}({\bf r})\neq n_{_{\rm NL}}^{-}({\bf r})}$. The different distribution of the refractive indices [Fig.~\ref{Fig_hybrid}(b) and \ref{Fig_hybrid}(c)] leads to a magneto-electric coupling~[see Ref~\cite{Alaee:15D} for a magneto-electric coupling in a reciprocal antenna]. To understand the underlying physics of the asymmetric nonreciprocal response and magneto-electric coupling, we compute the induced multipole moments using the exact multipole expansion for the opposite illuminations~[see Fig.~\ref{Fig_hybrid_radiation}(a) and ~\ref{Fig_hybrid_radiation}(b)]. The induced multipole moments are significantly different for the top and bottom illuminations which lead to intensity-dependent magneto-electric coupling.


When the hybrid antenna is illuminated from the bottom at low intensities, the antenna exhibits only electric dipole (ED) and magnetic quadrupole (MQ) moments, as can be seen in Fig.~\ref{Fig_hybrid_radiation}(a). The scattered electric fields of the ED and MQ moments are in-phase as they have even parities, and thus, interfere constructively in both forward and backward directions [see Eq.~(\ref{Csca_FB})]. Therefore, the hybrid antenna exhibits identical backward and forward scattering cross sections for the bottom illumination at low intensities: $C_{\mathrm{sca}}^{\mathrm{F}}=C_{\mathrm{sca}}^{\mathrm{B}}=C_{\mathrm{norm}}\left|p_{x}-ikQ_{yz}^{m}/6c\right|^{2}$, where $C_{\mathrm{norm}}$ is defined after Eq.~\ref{Csca_FB}. The radiation patterns in the forward and backward directions are also similar at low intensities~[see the inset of Fig.~\ref{Fig_hybrid_radiation}(c)]. By increasing the intensity, the contribution of
the ED and MQ moments in the scattering cross section remain constant, while the effect of the MD moment increases gradually~[see Fig. \ref{Fig_hybrid_radiation}(a)]. Therefore, at high laser intensities, e.g., $I_0=100\,\rm{GW/cm^2}$, the induced ED, MD, and MQ  moments interfere constructively (destructively) in the forward (backward) direction. Consequently, the antenna exhibits a nearly unidirectional radiation pattern with very small backscattering~[see Fig.~\ref{Fig_hybrid_radiation}(c)], a phenomenon known as the generalized Kerker effect~\cite{Alaee:2015generalized,Liu:2018generalized}. 

When the hybrid antenna is illuminated from the top, compared to the bottom illumination, different multipole moments contribute to the scattering~[see Fig.~\ref{Fig_hybrid_radiation}(a) and \ref{Fig_hybrid_radiation}(b)]. However, similar to the bottom illumination, the backward (forward) scattering decreases (increases) with the intensity, as shown in Fig.~\ref{Fig_hybrid_radiation}(d). At  $I_0\approx25\,\rm{GW/cm^2}$, the forward and backward scattering become identical~[see Fig.~\ref{Fig_hybrid_radiation} (d)].
Large tunability of the induced multipole moments of the hybrid antenna by an intense pump beam allows to control the radiation patterns from backward to forward directions and vice versa~[compare three radiation patterns in the inset of Fig.~\ref{Fig_hybrid_radiation}(d) corresponding to different intensities]. Therefore, by employing an \textit{ultrafast} optical pump~\cite{Alam:2016}, the radiation pattern of the hybrid antenna can be switched from a non-directive radiation to a directive one within a subpicosecond timescale~[see inset of Fig.~\ref{Fig_hybrid_radiation}(c) and \ref{Fig_hybrid_radiation}(d)]. Moreover, the hybrid antenna exhibits nonreciprocal radiation patterns because of nonreciprocal magneto-electric coupling.

In summary, we have theoretically studied nonlinear antennas based on ENZ materials with tunable absorption and scattering cross sections, as well as radiation patterns. We incorporated the extremely large and ultrafast nonlinear response of ENZ materials, in particular, ITO. We showed that while the radiation pattern of a single ITO antenna remains insensitive to the laser intensity, its absorption, scattering, and extinction cross sections can be modulated dynamically. Therefore, the antenna can be tuned between superabsorbing and superscattering states by controlling the intensity of the laser. Furthermore, we proposed a hybrid antenna composed of ITO and silicon disks with a radiation pattern that can be tuned between bidirectional and unidirectional emission under ultrafast optical pumping. Moreover, we found that the hybrid nanoantenna exhibits tunable nonreciprocal radiation patterns when illuminated from opposite directions because of the broken spatial inversion symmetry and large optical nonlinearity of ITO. We explained our findings based on the interference among the induced \textit{intensity-dependent} electric and magnetic multipole moments. The proposed tunable hybrid antenna with magneto-electric response can be used as a building block to design \textit{ultrafast switchable nonreciprocal} electric and magnetic mirrors~\cite{Asadchy:2015,Asadchy:2020}, metalens~\cite{Asadchy:2020}, metaabsorbers~\cite{Radi:2015Absorber,Alaee:2017Review}, metagrating~\cite{Radi:2015} and photonic topological insulators~\cite{Slobozhanyuk:2017}. In addition, our work provides a novel approach for designing tunable nanoantennas based on ENZ materials with a large optical nonlinearity.

\textbf{Acknowledgments.—}  R.A. acknowledges the support of the  Alexander von Humboldt Foundation through the Feodor Lynen Fellowship. L.C. acknowledges the support of China Scholarship Council. L.C., R.A., A.S., M. K., and R.W.B. are grateful to Boris Braverman and Jeremy Upham for helpful discussions and acknowledge support through the Natural Sciences and Engineering Research Council of Canada, the Canada Research Chairs program, and the Canada First Research Excellence Fund. R.W.B. also acknowledges support from the US Army Research Office and the US Defense Advanced Research Projects Administration.\\
L.C. and R.A. equally contributed to the work.

\begin{widetext}

\section{Multipole Expansion in Cartesian coordinates}

In this section, we present expressions for the exact induced multipole moments in Cartesian coordinates and the corresponding scattering and extinction cross sections~\cite{Alaee:2018,Alaee2019}. Having the induced nonlinear (intensity-dependent) polarization current, i.e. $\mathbf{J}_{\rm NL}\left(\mathbf{r},\omega\right)=-i\omega\left[\varepsilon_{\rm NL}\left(\mathbf{r},\omega\right)-\varepsilon_{0}\right]\mathbf{E}\left(\mathbf{r},\omega\right)$, the induced multipole moments can be obtained by using \textit{exact} multipole expansion introduced by~\citet{Alaee:2018,Alaee2019}:
\begin{eqnarray}
p_{\alpha} &=&-\frac{1}{i\omega}\left\{ \int d^{3}\mathbf{r}J_{\alpha, \rm NL}\,j_{0}\left(kr\right)+\frac{k^{2}}{2}\int d^{3}\mathbf{r}\left[3\left(\mathbf{r}\cdot\mathbf{J}_{\omega}\right)r_{\alpha}-r^{2}J_{\alpha, \rm NL}\right]\frac{j_{2}\left(kr\right)}{\left(kr\right)^{2}}\right\},\nonumber\\
m_{\alpha}&=&\frac{3}{2}\int d^{3}\mathbf{r}\left(\mathbf{r}\times\mathbf{J_{\rm NL}}\right)_{\alpha}\frac{j_{1}\left(kr\right)}{kr},\nonumber\\
Q_{\alpha\beta}^{\mathrm{e}} & = & -\frac{3}{i\omega}\left\{ \int d^{3}\mathbf{r}\left[3\left(r_{\beta}J_{\alpha,\rm {NL}}+r_{\alpha}J_{\beta,\rm{NL}}\right)-2\left(\mathbf{r}\cdot\mathbf{J}_{\rm{NL}}\right)\delta_{\alpha\beta}\right]\frac{j_{1}\left(kr\right)}{kr}\right.\nonumber\\
 &  & \left.+2k^{2}\int d^{3}\mathbf{r}\left[5r_{\alpha}r_{\beta}\left(\mathbf{r}\cdot\mathbf{J}_{\rm NL}\right)-\left(r_{\alpha}J_{\beta, \rm {NL}}+r_{\beta}J_{\alpha, \rm {NL}}\right)r^{2}-r^{2}\left(\mathbf{r}\cdot\mathbf{J}_{\rm NL}\right)\delta_{\alpha\beta}\right]\frac{j_{3}\left(kr\right)}{\left(kr\right)^{3}}\right\},\nonumber\\
 Q_{\alpha\beta}^{m}&=&15\int d^{3}\mathbf{r}\left\{ r_{\alpha}\left(\mathbf{r}\times\mathbf{J_{\rm NL}}\right)_{\beta}+r_{\beta}\left(\mathbf{r}\times\mathbf{J_{\rm NL}}\right)_{\alpha}\right\} \frac{j_{2}\left(kr\right)}{\left(kr\right)^{2}}, \label{eq:ME}
\end{eqnarray}
where, $\alpha,\beta \in {x,y,z}$, and where, $p_{\alpha}$, $m_{\alpha}$, $Q_{\alpha\beta}^e$, and $Q_{\alpha\beta}^m$ are the
electric dipole (ED), magnetic dipole (MD), electric quadrupole (EQ), and magnetic quadrupole (MQ) multipole moments, respectively. $j_{n}\left(kr\right)$ is the nth spherical Bessel function. 

The total scattering cross section of the nonlinear antenna (sum of the induced multipole moments) is given by~\cite{Alaee:2018}
\begin{eqnarray}\label{Csca}
C_{\rm sca}&=&\frac{k^4}{6\pi{\varepsilon_0}^2|E_0|^2}\left[\sum\limits_{\alpha}(|p_{\alpha}|^2+|\frac{m_{\alpha}}{c}|^2)
+\frac{1}{120}\sum\limits_{\alpha\beta}(|kQ_{\alpha\beta}^e|^2+|\frac{kQ_{\alpha\beta}^m}{c}|^2)\right],
\end{eqnarray}
and the extinction cross section reads as
\begin{eqnarray}
C_{\mathrm{ext}} & = & \frac{k}{\varepsilon_{0}E_{0}^{2}}\mathrm{Im}\left\{ \sum_{\alpha=1}^{3}\left(p_{\alpha}E_{\alpha,{\rm inc}}^{*}+\frac{m_{\alpha}}{c}Z_{0}H_{\alpha,{\rm inc}}^{*}\right)\right.\nonumber\\
 &  & \left.\sum_{\alpha,\beta=1}^{3}\left[Q_{\alpha,\beta}^{e}\left(\frac{\partial E_{\beta,\mathrm{inc}}^{*}}{\partial x_{\alpha}}+\frac{\partial E_{\alpha,\mathrm{inc}}^{*}}{\partial x_{\beta}}\right)+\frac{Q_{\alpha,\beta}^{m}}{c}Z_{0}\left(\frac{\partial H_{\beta,\mathrm{inc}}^{*}}{\partial x_{\alpha}}+\frac{\partial H_{\alpha,\mathrm{inc}}^{*}}{\partial x_{\beta}}\right)\right]\right\} 
\end{eqnarray}
where $E_0$ is the amplitude of incident field. ${\mathbf E}_{\rm inc}$, and ${\mathbf H}_{\rm inc}$ are the incident electric and magnetic fields, respectively. $Z_0$ is the impedance of the free space and $c$ is the speed of the light in free space.

\begin{figure}
\begin{centering}
\includegraphics[width=9cm]{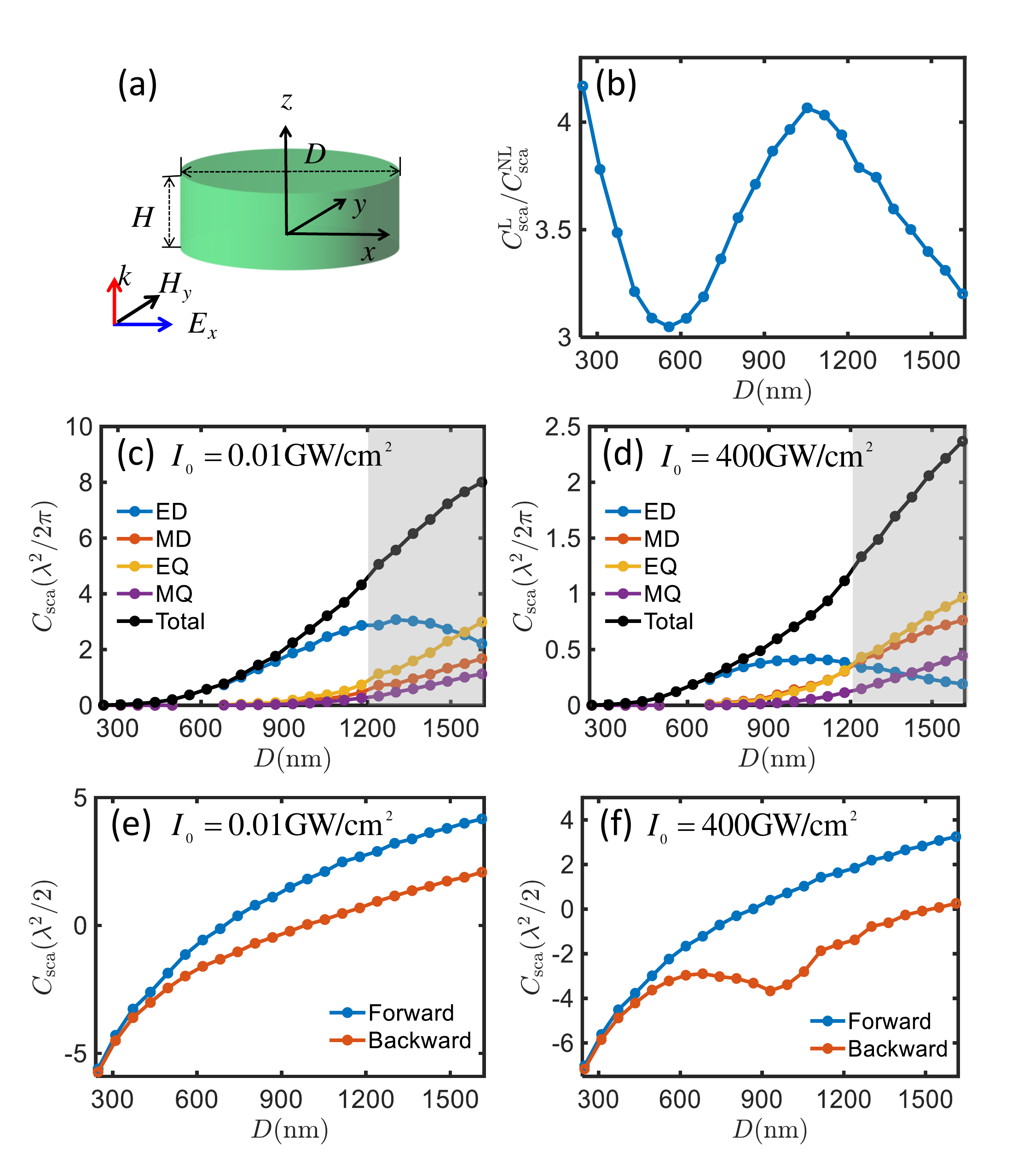}
\par\end{centering}
\caption{(a) Schematic drawing of the ITO disk with height $H$ and diameter $D$. (b) The ratio of the linear ($I_0=0.01$ $\mathrm{GW}/\mathrm{cm^2}$) to the nonlinear ($I_0=400$ $\mathrm{GW}/\mathrm{cm^2}$) scattering cross section. (c) and (d) Total scattering cross sections (normalized to $\lambda^2/2\pi$) and contributions of different electric and magnetic multipole moments: electric dipole (ED), magnetic dipole (MD), electric quadrupole (EQ), magnetic quadrupole (MQ) in low and high intensities, respectively. (e) and (f) Normalized forward and backward scattering cross section for low and high intensities, respectively.
\label{Fig. S1}}
\end{figure}
\begin{figure}
\begin{centering}
\includegraphics[width=9cm]{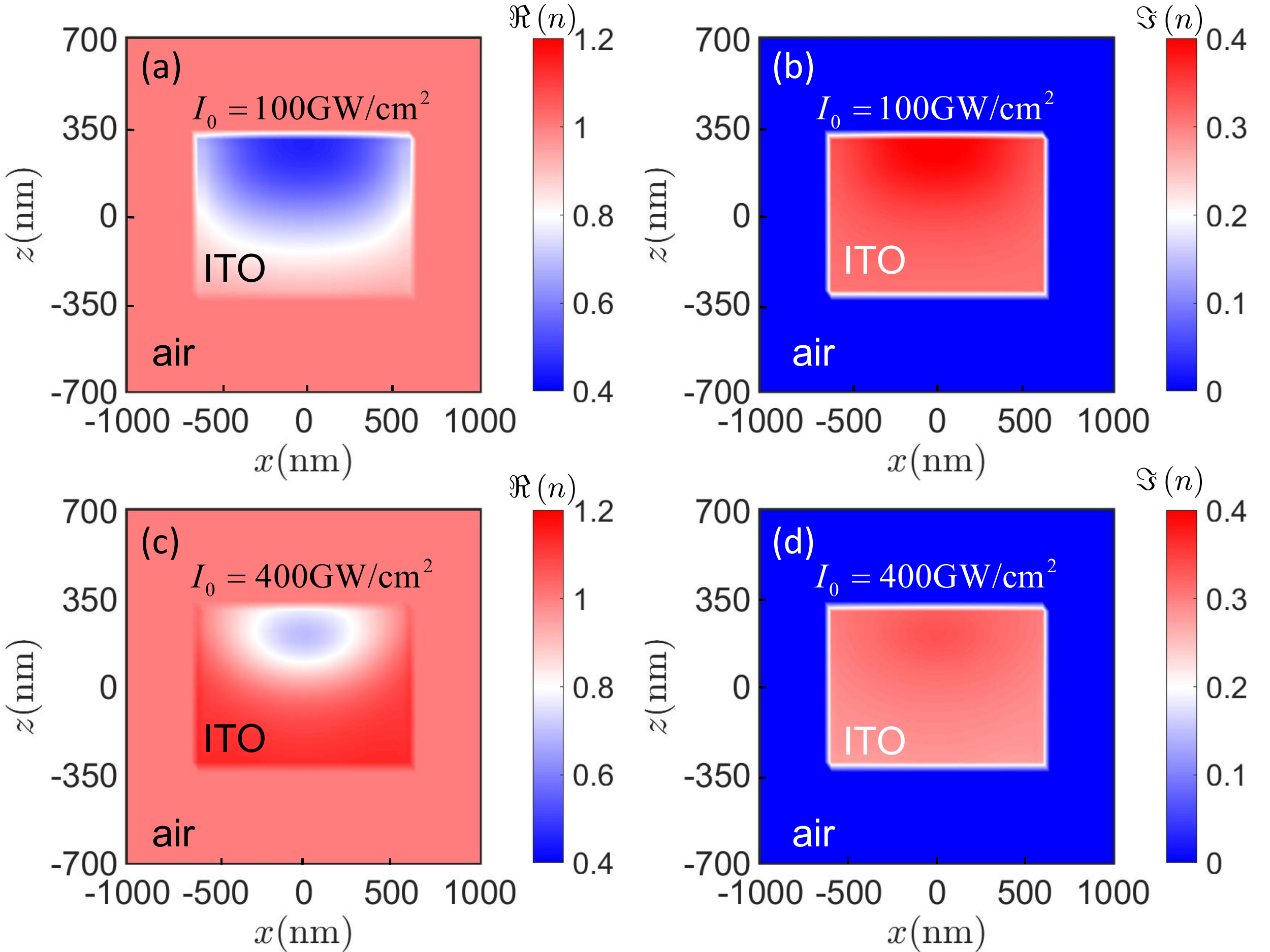}
\par\end{centering}
\caption{(a)-(b) Real and imaginary parts of the refractive index in the $xz$-plane (at $y=0$) at input intensity $I_0=100$ $\mathrm{GW}/\mathrm{cm^2}$. (c)-(d) Same as (a)-(b) for laser intensity $I_0=400$ $\mathrm{GW}/\mathrm{cm^2}$. The dimensions of the ITO disk are $D=2H=1200$\,nm.}
\label{Fig. S2}
\end{figure}
\section{Far Fields and Radiation Patterns}
Using the induced multipole moments in Eq.~\ref{eq:ME}, the far field can be found ~\cite{Jackson1999,Alaee:2015generalized}
\begin{eqnarray}
\mathbf{E}_{\rm ED} & = & \frac{k^{2}}{4\pi\epsilon_{0}}\frac{e^{ikr}}{r}p_{x}\left(-\mathrm{sin}\varphi\mathbf{e}_{\varphi}+\mathrm{cos}\theta\mathrm{cos}\varphi\mathbf{e}_{\theta}\right),\nonumber \\
\mathbf{E}_{\rm MD} & = & \frac{k^{2}}{4\pi\epsilon_{0}}\frac{e^{ikr}}{r}\frac{m_{y}}{c}\left(-\mathrm{cos}\theta\mathrm{sin}\varphi\mathbf{e}_{\varphi}+\mathrm{cos}\varphi\mathbf{e}_{\theta}\right),\nonumber \\
\mathbf{E}_{\rm EQ} & = & \frac{k^{2}}{4\pi\epsilon_{0}}\frac{e^{ikr}}{r}\frac{ik}{6}Q_{zx}^{e}\left[\mathrm{cos}\theta\mathrm{sin}\varphi\mathbf{e}_{\varphi}-\mathrm{\left(\mathrm{2cos^{2}}\theta-1\right)cos}\varphi\mathbf{e}_{\theta}\right],\nonumber \\
\mathbf{E}_{\rm MQ} & = & \frac{k^{2}}{4\pi\epsilon_{0}}\frac{e^{ikr}}{r}\frac{ik}{6c}Q_{zy}^{m}\left[\left(\mathrm{2cos^{2}}\theta-1\right)\mathrm{sin}\varphi\mathbf{e}_{\varphi}-\mathrm{\mathrm{cos}\theta cos}\varphi\mathbf{e}_{\theta}\right],\label{eq:E_Far}
\end{eqnarray}
where $r,\theta,\varphi$ are the radial distance, polar angle, and
azimuthal angle, respectively. Considering the contribution from all multipole moments (up to magnetic quadrupole), the electric field corresponding to the radiation pattern ($\propto\left|\mathbf{E}\right|^{2}$) in the $xz$-plane, i.e., $\varphi=0$, can be written as
\begin{eqnarray}
\mathbf{E} & \approx & \frac{k^{2}}{4\pi\epsilon_{0}}\frac{e^{ikr}}{r}\left[p_{x}\mathrm{cos}\theta+\frac{m_{y}}{c}-\frac{ik}{6}Q_{xz}^{e}\left(\mathrm{2cos^{2}}\theta-1\right)-\frac{ik}{6c}Q_{zy}^{m}\mathrm{cos}\theta\right]\mathbf{e}_{\theta},\label{E_rad}
\end{eqnarray}
from which, the radiation pattern ($\propto\left|\mathbf{E}\right|^{2}$) can be obtained. Equations~(\ref{eq:E_Far}) and (\ref{E_rad}) are used to plot the radiation patterns in the main text.

\begin{figure}
\begin{centering}
\includegraphics[width=9cm]{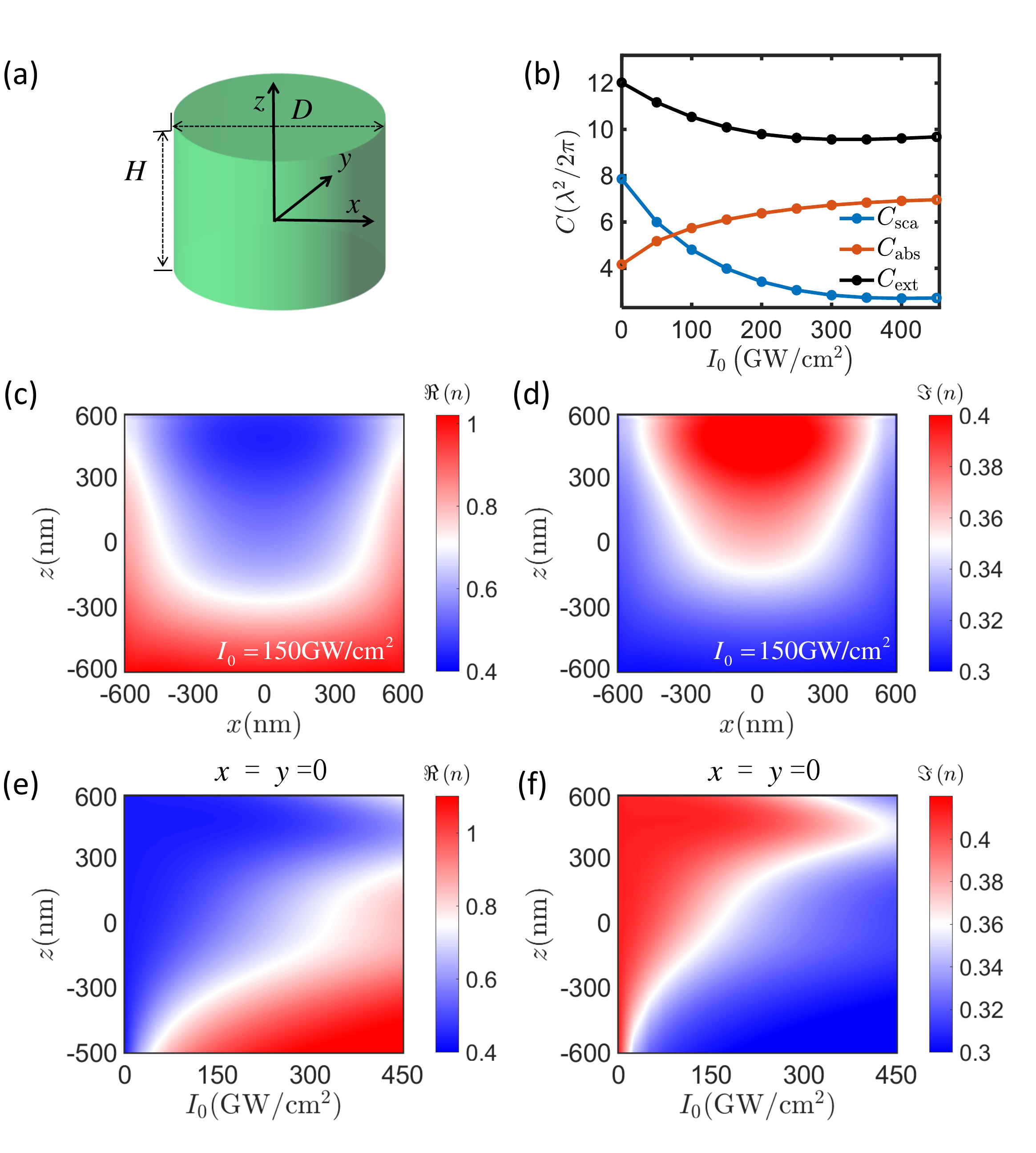}
\par\end{centering}
\caption{(a) Schematic drawing of the ITO disk with height $H$ and diameter $D$. (b) Scattering, absorption and extinction cross sections as a function of intensity $I_0$ for $H=D=1200$\,nm. (c) and (d) The real and imaginary part of the refractive index in the $xz$-plane (at y=0) at laser intensity $I_0=150$ $\mathrm{GW}/\mathrm{cm^2}$. (e) and (f) The real and imaginary parts of the refractive index along the $z$-axis at $x=y$=0 as a function of the laser intensity.}
\label{Fig. S3}
\end{figure}
\section{Nonlinear antennas based on ENZ materials}
In this section, we discuss the intensity-dependant refractive index and scattering cross sections of the ITO antenna.

\subsection{Intensity-dependant scattering cross sections}
Figure~\ref{Fig. S1}(a) shows a schematic drawing of the ITO disk when illuminated by an $x$-polarized plane wave propagating in the $z$ direction. Figure~\ref{Fig. S1}(b) shows the ratio of the linear (low intensity, i.e., $I_0=0.01$ $\mathrm{GW}/\mathrm{cm^2}$) to the nonlinear (high intensity, i.e., $I_0=400$ $\mathrm{GW}/\mathrm{cm^2}$) scattering cross section. It can be seen that, using the strong nonlinear response of ITO, the scattering cross section of the antenna can be changed by more than a factor of three. Figure~\ref{Fig. S1}(c) and (d) show the total scattering cross sections (normalized to $\lambda^2/2\pi$) and the contributions of the electric and magnetic multipole moments of the ITO antennas as a function of the diameter of the ITO disk. For small diameters, $D<$\,600nm, the antenna exhibits only an electric dipole moment and the higher-order multipole moments are very small~[See Fig.~\ref{Fig. S1}(c) and \ref{Fig. S1}(d)]. Thus, the antenna exhibits an omnidirectional radiation pattern at both low ($I_0=0.01$ $\mathrm{GW}/\mathrm{cm^2}$) and high ($I_0=400$ $\mathrm{GW}/\mathrm{cm^2}$) intensities. Figure~\ref{Fig. S1} (e) and (f) depicts the forward and backward scattering cross sections. For small diameters, we observe an identical forward and backward scattering cross section which indicates an omnidirectional radiation. At larger diameters $D$, the ITO antenna supports higher order multipoles in addition to ED~[see Fig. \ref{Fig. S1}(c)-(d)]. As a result, the forward and backward scatterings are significantly different~[see e.g., $D>900$\,nm in Fig. \ref{Fig. S1}(e)-(f)]. 

\subsection{Intensity-dependant refractive index}
In the main text, we plot only the real part of the refractive index in Fig. 1. Here, we show the real and imaginary parts of the intensity-dependent refractive index of ITO at the ENZ wavelength $\lambda_{\rm ENZ}=1240$\,nm for an ITO disk~( Fig.~\ref{Fig. S2}). In our simulations, we used Maxwell’s equations numerical solver combined with an iterative technique to solve for the intensity-dependent refractive index. The refractive index depends on position $\mathbf{r}$ because of the induced \textit{nonuniform} electric field distribution inside the ITO antenna. At very low intensities, the refractive index is uniform inside the ITO disk~[not shown here]. Whereas, at high intensities~[Fig.~\ref{Fig. S2}(a) and (d)], the refractive index changes significantly due to the large Kerr-type nonlinear response of ITO. 

\section{Superabsorption and superscattering}
In this section, we show the intensity-dependent refractive index and multipole moments for the superabsorption and superscattering states. 

\subsection{Intensity-dependant refractive index}

Figure~\ref{Fig. S3}(b) plots the scattering, absorption and extinction cross sections as functions of intensity for $H = D = 1200$\,nm  for superabsorption and superscattering cases~[same as Fig. 3 in the main text for the ITO disk shown in  Fig.~\ref{Fig. S3}(a)]. The real part of the refractive index in the $xz$-plane (at $y=0$) at intensity $I_0=150$ $\mathrm{GW}/\mathrm{cm^2}$ is shown in Fig.~\ref{Fig. S3}(c)-(d). We observe a position-dependent refractive index. We also plot the real and imaginary parts of the intensity-dependent refractive index in Fig.~\ref{Fig. S3}(e)-(f) as a function of $z$ at $x=y=0$.

\begin{figure}
\begin{centering}
\includegraphics[width=9cm]{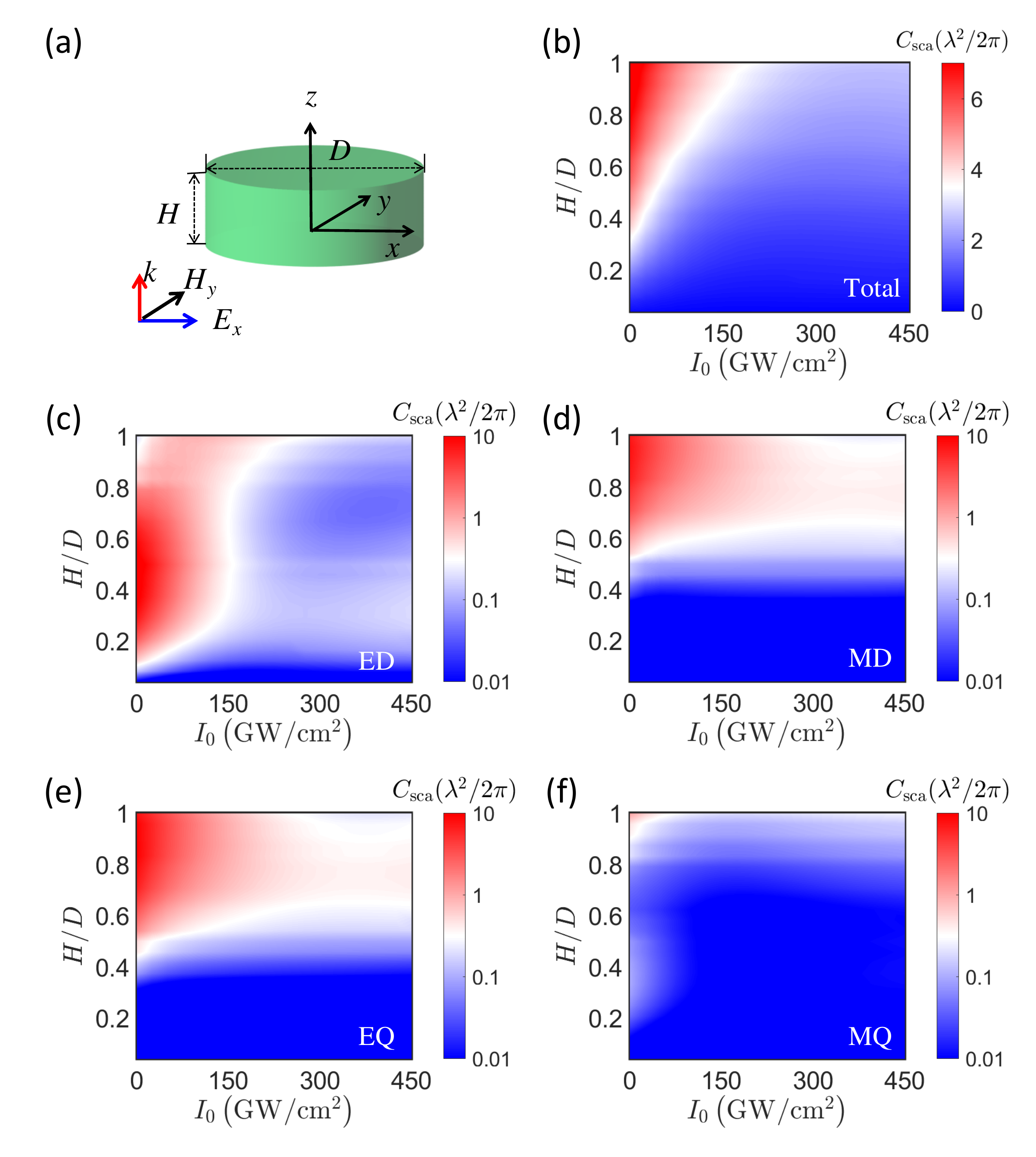}
\par\end{centering}
\caption{(a) Schematic drawing of the ITO disk with height $H$ and diameter $D$. (b) Scattering cross section $C_{\rm sca}$ (normalized to $\lambda^2/2\pi$) of the ITO antenna as a function of normalized height, i.e., $H/D$ and intensity of laser $I_0$, where the diameter of the antenna $D=1200$\,nm. (c)-(f) contribution of different multipole moments ED, MD, EQ and MQ, respectively, in the total scattering cross section shown in (b).}
\label{Fig. S4}
\end{figure}
\subsection{Intensity-dependant electric and magnetic multipole moments}

In the main text, we show that the absorption and scattering cross sections of the ITO antenna can be controlled from a nearly superscatterer to a nearly superabsorber by changing the intensity of the laser beam. This results can be understood in terms of the Intensity-dependant electric and magnetic multipole moments. Figure~\ref{Fig. S4}(b) shows the total scattering cross section of the ITO disk drawn in Fig.~\ref{Fig. S4}(a) as a function of intensity. The contributions of the electric and magnetic multipole moments as a function of height-to-diameter ratio, $H/D$, and the laser intensity $I_0$ are shown in Fig.~\ref{Fig. S4}(c)-(f). It can be seen that by changing the intensity of the laser beam, we can significantly control the contribution of electric and magnetic multipole moments.

\end{widetext}


%

\end{document}